# Electrochemical Impedance Spectroscopy of a novel ZnA-CA (Zinc Acetate – Citric Acid) Supramolecular Metallogel

**Saranya Babu[1#], Aiswarya M[1], Subhendu Dhibar[2], Goutam Chandra[1], P.Predeep[3,1*]**

[1]*Laboratory for Molecular Photonics and Electronics (LAMP), Department of Physics, National Institute of Technology Calicut, Kozhikode.*

[2]*Department of Chemistry, The University of Burdwan, Golapbag, Burdwan-713104*

[1,3]*School of Nanoscience and Nanotechnology, Mahatma Gandhi University, Kottayam, Kerala, India*

[#] *TK Madhava Memorial College, Nangiarkulangara, Kerala, India*

## Abstract

Supramolecular gels are formed by the self-assembly of low molecular weight organic gelators with their solvent molecules. These are emerging novel materials with good semiconducting and light emitting properties with application potential as hole and electron transport layers in organic solar cells, LEDs, stimulus-responsive smart semiconducting materials, thin film transistors (TFT) etc. In this context charge transport and mobility of charge carriers in these materials assume extreme significance. In the study, Electrochemical impedance spectroscopy, which is a non-destructive technique, is used to analyze frequency dependent electrochemical impedance values, of a novel meallogel, ZnA-CA (Zinc Acetate – Citric Acid), and used to evaluate the charge properties and mobility. A comparative study of mobility values obtained from diode I-V characteristics of the gel and Impedance measurements has also been made.

## Introduction

Electrochemical impedance spectroscopy is an electrical technique widely used for the characterization of organic photovoltaic devices. Using this non-destructive technique, we can probe into the charge transport properties of the fabricated devices. We can extract carrier mobility [1] of the charge carrier within the device and analyse the charge transport properties using this impedance spectroscopy (IS). This characterisation is based on a small perturbation [2] of the applied voltage. A constant DC voltage applied initially establishes a stationary situation inside the device, which is perturbed by a small AC signal of varying frequencies. IS has a simple experimental set up which is quite easy to handle and this technique facilitates the extraction of different physical quantities at the same time.

**Material and Device Fabrication**

The ZnA- CA metallogel is to be sandwiched between the anode and cathode for impedance spectroscopy measurements. This is accomplished by spin coating the material on an ITO followed by thermal annealing for 5 minutes to remove excess solvent. Metal electrode is deposited over this by vacuum evaporation technique. ITO serves as the anode which establishes an ohmic contact with the material and the non–injection contour electrode is Al electrode (cathode) [3]. The prepatterned ITO substrates are cleaned by the standard cleaning procedure which includes sonication in deionised water, isopropyl alcohol and 15-minute long uv- ozone exposure. The metallogel- PMMA blend in DMF is spin coated over this ITO substrate at spin speed of 1500 rpm for 60s. These films having a thickness range of 200-300 nm are thermally annealed at 60 $^0$C for 5 minutes. 100 nm thick Aluminium (Al) electrodes are deposited over this film by vacuum evaporation technique. Diode type device structures of the form, ITO/ Gel-PMMA blend/Al (Fig.1) with effective area 0.04 cm$^2$ is obtained using a shadow mask. Impedance spectroscopy was performed with Wayne Kerr impedance analyser with 25 mV amplitude and frequency ranging from 20 Hz to 10 M Hz.

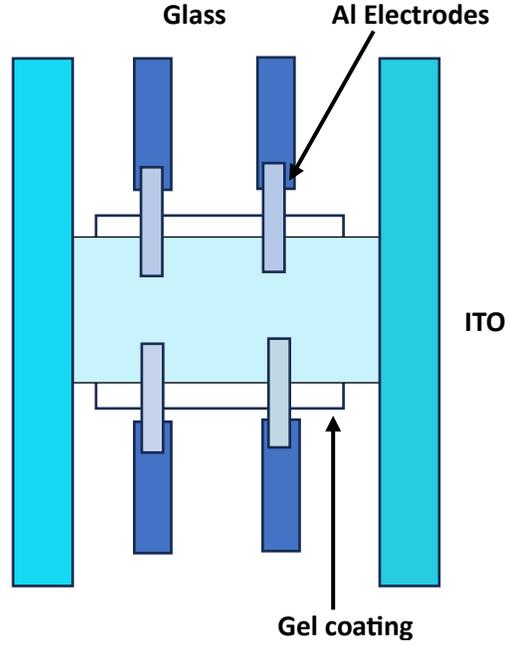

**Fig. 1**: ITO/ Gel + PMMA / Al device structure analysis

**Theory**

In impedance spectroscopy measurement, a sinusoidally varying voltage $V_{ac}$ is the applied to the sample giving a sinusoidally varying current $I_{ac}$ as the resultant [1]. In time domain, they are denoted by,[3];

$$V(t) = V_0 \sin \omega t = V_0 e^{i\omega t} \qquad (1)$$

$$I(t) = I_0 \sin(\omega t + \phi) = I_0 e^{i(\omega t + \phi)} \qquad (2)$$

Where, $\omega$ is the angular frequency ($\omega = 2\pi f$), $\phi$ is the phase delay between $V(t)$ and $I(t)$, $V_0$ and $I_0$ are the amplitude of the voltage and current signals, respectively and t is time. The impedance Z(t) is given by

$$Z(t) = \frac{V(t)}{I(t)} = Z_0 e^{-i\phi} \qquad (3)$$

Z is a transfer function that relates time – varying voltage ($V_{ac}$) and current ($I_{ac}$) as a function of frequency. The ratio of $V_{ac} : I_{ac}$ denotes the frequency dependent or time-varying complex resistance of the sample. By varying the frequency, the input voltage from 20 Hz to 10 MHz we get a spectrum of impedance values. In complex plane it can be represented as

$$Z = Z' - iZ'' \qquad (4)$$

where, $Z'$ is the real part and $Z''$ is the imaginary part of $Z$. And the magnitude of complex impedance is;

$$Z_0 = \sqrt{(Z'^2 + Z''^2)}. \qquad (5)$$

By Ohm's law, the ratio of voltage and current is resistance, the hinderance to the current flow in the sample. The real part of impedance $(Z')$ is equivalent to resistance R, and the imaginary part $(Z'')$ equals to the reactance.

$$Z' = R = Z_0 \cos(-\phi) = Z_0 \cos\phi \qquad (6)$$

$$Z'' = X = Z_0 \sin(-\phi) = Z_0 \sin\phi \qquad (7)$$

From this, the complex admittance; $Y(\omega)$ is written as,

$$Y(\omega) = G + iB = G + i\omega C \qquad (8)$$

Where, G is conductance, $\omega$ is angular frequency, B is susceptance and C is equivalent capacitance of the device. The imaginary part of this expression gives the frequency dependent capacitance.

When AC bias is given to the sample, capacitance C changes according to the frequency. At high frequencies, (f > $\tau_r^{-1}$), C becomes constant since at that stage it is bias dependent and only the dielectric effects persists [3]. Capacitance is equal to geometric capacitance ($C_{geo}$) at this stage. At low frequencies, capacitance increase due to slow trapping effects [3][4]. At high frequencies, due injected carriers the phase difference between $V_{ac}$ and $I_{ac}$ decreases and so C decreases.

Using the capacitance values obtained from impedance measurements and the $C_{geo,}$ negative differential susceptance ($-\Delta B$) is computed from the equation 9 [5].

$$-\Delta B = -2\pi f[C(f) - C_{geo}] \qquad (9)$$

Where, $C(f)$ is the capacitance of the device at frequency $f$ and $C_{geo}$ is the geometrical capacitance. The frequency- negative differential susceptance plot shows a maximum at the frequency value, $f_{max} = \tau_r^{-1}$, where $\tau_r$ is correlated to mean transit time ($\tau_t$) by [6];

$$\tau_t = 0.72\tau_r$$
$$(10)$$

Transit time is connected to charge carrier mobility ($\mu$) by [6],

$$\mu = \frac{4}{3}\frac{d^2}{\tau_t V} \quad (11)$$

Here, $V$ is the voltage applied to the film of thickness $d$. Here, transit time and mobility are determined from frequency dependence of negative differential susceptance.

**Result and Discussion**

**Determination of charge carrier mobility from the frequency response of capacitance and negative differential susceptance (-ΔB) of ZnA – CA Metallogel**

The charge transport properties of ZnA-CA metallogel is analyzed using impedance spectroscopy measurements. Frequency dependent capacitance measurements are used for the extraction of charge carrier mobility. The device structure is same as in figure 1. The frequency response of the device is shown in figure 2.

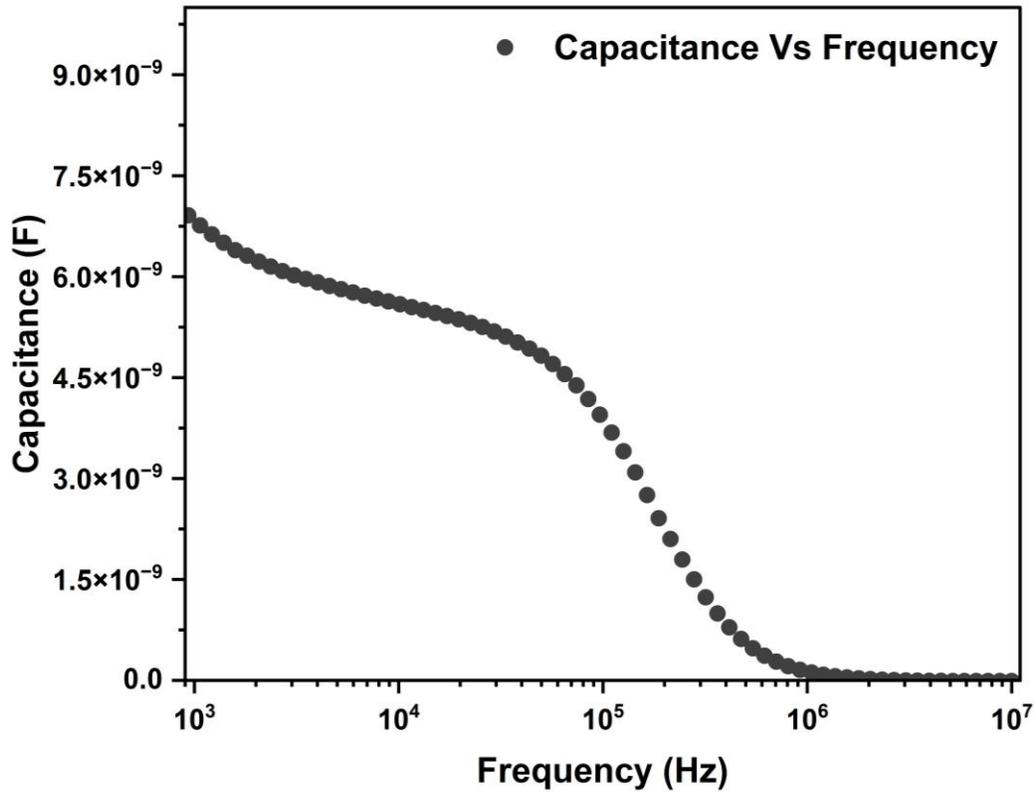

**Fig.2**. Frequency- Capacitance curves of ITO/ ZnA-CA Gel + PMMA / Al devices with

PMMA concentration 40 %.

Slow traps causes hike in capacitance values in the low frequency regions due to the trapping and de trapping effects [7]. The crystallinity variation in the devices modifies the trap density [8] thereby affecting conductivity. When the PMMA concentration is 40 % have high crystallinity. High percentage of crystallinity leads to improved charge carrier mobility [8]. The high capacitance at low frequency is due to the contribution from the trapped states [9]. At high frequencies, these trap states respond slowly and capacitance fall of exponentially and finally saturates. The frequency verses negative differential susceptibility of the device is shown in figure 3.

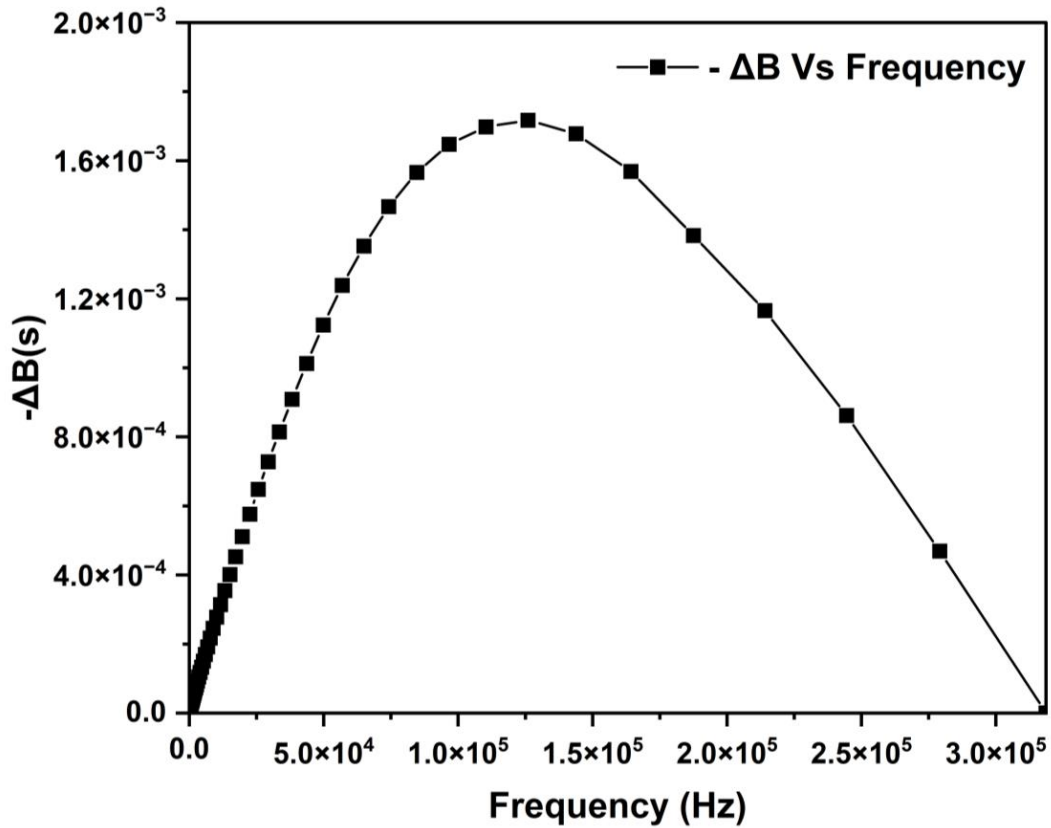

**Fig.3**. Frequency- negative differential susceptance curves of curves of ITO/ Gel + PMMA / Al devices with PMMA concentration 40%.

The frequency vs. capacitance values obtained from impedance spectra are used to calculate negative differential susceptance (-$\Delta B$). Addition of PMMA for film preparation ensure uniform film formation which favours smooth drifting of charge carriers. The reduction in transit time helps in reducing charge trapping and de trapping. Uniform thin films devoid of pin holes is the mandatory criteria for better device performance. PMMA acts as a binder which inhibits coagulation[10] and helps in the formation smooth uniform thin films with negligible pin holes. This avoids charge trapping, thereby facilitating smooth charge drifting. The crystallinity variation in the devices modifies the trap density thereby affecting conductivity. High percentage of crystallinity leads to improved charge carrier mobility ($\mu = 3.819 \times 10^{-2}$ cm$^2$/Vs). The combined effect of high crystallinity and film quality give positive impetus to the charge pathways. FTIR and Raman analysis confirm that the basic nature of the bonds and chemical environment remains unaltered[10] even after PMMA addition.

**Conclusion**

Charge carrier mobility of the Schottky diodes with ZnA -CA metallogel employing PMMA as binder is done. The frequency dependance of the slow traps shows gradual changes with enhancement in PMMA concentration. For ZnA-CA based devices, the device with 40 % PMMA offers exceptionally high carrier mobility also. The mobility values obtained from I-V characteristics and Impedance measurements for both metallogels are of the same order. These studies confirm the usefulness of materials under study for semiconducting applications.